# Engineering Topological Materials

Amit Goft[1] and Eric Akkermans[1]

[1]*Department of Physics, Technion – Israel Institute of Technology, Haifa 3200003, Israel*

The tenfold classification provides a powerful framework for organizing topological phases of matter based on symmetry and spatial dimension. However, it does not offer a systematic method for transitioning between classes or engineering materials to realize desired topological properties. In this work, we introduce a general method for designing topological materials by embedding defects or spatial textures, which alter symmetry or dimension. This enables controlled navigation across the tenfold table, allowing one to induce topological phase transitions on demand. We illustrate this approach through several nontrivial examples, demonstrating how local defects can generate phases with different symmetries and topological invariants.

Interest in topological phases was sparked by the quantum Hall effect and the classification of Bravais lattices [1, 2], ultimately leading to the discovery of topological quantum materials [3, 4]. The tenfold classification of band insulators and superconductors [5] provides a systematic framework for identifying topological features in weakly interacting fermions in spatial dimension $d$. It organizes families of Hamiltonians $H(d, s)$ according to the existence of two antiunitary symmetries: time-reversal symmetry (T) and particle-hole symmetry (P). Each of them can appear in one of three distinct forms: absent ($S = 0$) or present with $S^2 = \pm 1$. A third, the chiral symmetry (C), is essentially defined as the product of the two antiunitary symmetries, $C = PT$. C is unitary and satisfies the anticommutation relation $\{C, H(d, s)\} = 0$ whenever it exists. This leads to ten distinct symmetry classes, denoted by the triple $(T, P, C)$, each labeled by a symmetry index $s$ [5]. These classes are organized into the tenfold table of topological insulators and superconductors [6, 7], which categorizes topological phases as a function of $(s, d)$, as shown in Table I. The table was further generalized [8, 9] to include defects or textures by introducing the codimension $\delta = d - D$, which effectively lowers the spatial dimension of the system. Despite significant progress, challenges remain: the tenfold classification allows for the identification and classification of topological materials, but it does not provide a mechanism for how to design or induce on-demand topological phases. In addition, the tenfold classification is plagued by its underlying, hard-to-decipher mathematical structure, making it hard to efficiently leverage towards predicting topological phases.

This paper strives to address the current situation. We introduce a comprehensive approach for navigating the tenfold classification framework, which employs the symbol of a Hamiltonian [9], by introducing on-demand properly designed defects or textures. This enables the controlled construction of topological materials with prescribed symmetry $s$, reduced dimension $\delta$, and belonging to the topological class $\mathbb{Z}$ or $\mathbb{Z}_2$. This approach has broad implications. Practically, it allows for the transformation of non-topological phases into topological ones. Conceptually, it offers a comprehensive framework for generating both spinless and spin-1/2 excitations, encompassing fermions, bosons, and classical waves. For instance, our approach enables the construction of systems exhibiting effective $T^2 = -1$ symmetry, typical in spin one-half fermions, originating from systems characterized by $T^2 = +1$ symmetry. In addition, this framework establishes a link between various topological phases and clarifies attributes previously believed to be unique to fermionic systems, such as fermion doubling, demonstrating their wider relevance. Ultimately, it enables us to explore the connection between topology and quantum entanglement [10].

To illustrate this framework, we devise a method starting with systems that are translationally invariant and contain Dirac points. By adding spatially localized defects that couple low energy modes, we generate complex scalar fields that alter the topology while preserving chiral symmetry. The symmetry inherent in the underlying band structure determines the distinct classes, such as $BDI$ or $CII$, in which the resulting phase may reside (Table I). Analytical derivation of the conditions supporting these transitions allows for a precise engineering of topological phases.

In the tenfold classification, topological classes are defined by invariants which can either be integer ($\mathbb{Z}$) or binary ($\mathbb{Z}_2$). To calculate these invariants, one must use an extra mathematical object known as the symbol of a Hamiltonian. For a Hamiltonian $H(d, s)$, its symbol $\mathcal{H}(\boldsymbol{k}, \boldsymbol{r})$ is established through the discrete (Weyl) transform:

$$\mathcal{H}_{mn}(\boldsymbol{k}, \boldsymbol{r}) = \sum_j e^{-2i\boldsymbol{k}\cdot\boldsymbol{R}(j)} {}_m\langle \boldsymbol{r} + \boldsymbol{R}(j)| H |\boldsymbol{r} - \boldsymbol{R}(j)\rangle_n. \tag{1}$$

The symbol is a $w \times w$ matrix that depends on both the momentum $\boldsymbol{k}$ and the position $\boldsymbol{r}$. The indices $m, n$ correspond to internal (non-spatial) degrees of freedom, like sublattice, layer, valley, or spin. Essentially, the symbol preserves the complete internal structure of the system and treats spatial operators as parameters. It is important to highlight that this definition is specific to single-particle Hamiltonians, allowing the symbol to be assigned to all Hamiltonians within the tenfold classification. Even though the symbol generally does not capture the entire spectrum of the Hamiltonian $H$, it simplifies to the Bloch Hamiltonian $\mathcal{H}(\boldsymbol{k})$ in systems with translational symmetry, enabling direct determination of energy bands.

| Class | s | T | P | C | d=0 | 1 | 2 | 3 |
|-------|---|---|---|---|-----|---|---|---|
| A | 0 | 0 | 0 | 0 | $\mathbb{Z}$ | 0 | $\mathbb{Z}$ | 0 |
| AIII | 1 | 0 | 0 | 1 | 0 | $\mathbb{Z}$ | 0 | $\mathbb{Z}$ |
| AI | 0 | + | 0 | 0 | $\mathbb{Z}$ | 0 | 0 | 0 |
| BDI | 1 | + | + | 1 | $\mathbb{Z}_2$ | $\mathbb{Z}$ | 0 | 0 |
| D | 2 | 0 | + | 0 | $\mathbb{Z}_2$ | $\mathbb{Z}_2$ | $\mathbb{Z}$ | 0 |
| DIII | 3 | - | + | 1 | 0 | $\mathbb{Z}_2$ | $\mathbb{Z}_2$ | $\mathbb{Z}$ |
| AII | 4 | - | 0 | 0 | $2\mathbb{Z}$ | 0 | $\mathbb{Z}_2$ | $\mathbb{Z}_2$ |
| CII | 5 | - | - | 1 | 0 | $2\mathbb{Z}$ | 0 | $\mathbb{Z}_2$ |
| C | 6 | 0 | - | 0 | 0 | 0 | $2\mathbb{Z}$ | 0 |
| CI | 7 | + | - | 1 | 0 | 0 | 0 | $2\mathbb{Z}$ |

TABLE I. The tenfold table. The left five columns present the 10 symmetry classes identified by $s$, which are determined through their antiunitary symmetries $T, P$, and chirality $C$. The symbol "+" indicates that the corresponding operator is a symmetry that squares to 1, whereas "−" signifies squaring to −1, and "0" indicates its absence. The remaining four columns denote potential topological classes $(0, \mathbb{Z}, \mathbb{Z}_2)$ in relation to the dimension $d$. When considering defects, $d$ is replaced with the reduced dimension $\delta = d - D$. The table exhibits 8 periodicity both in symmetry class and dimension. We have limited the table to $d \leq 3$.

The symbol is crucial for identifying topological features, especially for elliptic differential operators [11–14]. These are operators whose symbol vanishes only at a limited set of isolated points. For instance, the Laplacian $\Delta = \sum_i \partial_i^2$ has the symbol $\mathcal{H}(\boldsymbol{k}) = \sum_i k_i^2$, which vanishes only at $\boldsymbol{k} = 0$. In lattice systems, comparable behavior appears in the continuum limit; for example, in a square lattice, the low-energy dispersion $E(\boldsymbol{k}) \sim \cos(k_x) + \cos(k_y) \approx \boldsymbol{k}^2$ also shows ellipticity.

Topological invariants within the tenfold classification scheme relate to the Hamiltonian's spectral characteristics through Atiyah–Singer Index Theorems (ASIT) [11, 13–24].These theorems establish a connection between the analytical index of an elliptic differential operator $\mathcal{Q}$ and a related topological invariant. The analytical index is given by Index $\mathcal{Q} = \dim \ker \mathcal{Q} - \dim \ker \mathcal{Q}^\dagger$, coinciding with a topological quantity given in the tenfold table. The zero modes of the operators $\mathcal{Q}$ and $\mathcal{Q}^\dagger$ are generally associated with eigenstates of the Hamiltonian. Consequently, the existence of these zero modes indicates a nontrivial topological signature in the spectrum of the Hamiltonian. A well-known instance in condensed matter physics is the integer quantum Hall effect, where the index is identified as a Chern number [25–27].

Classes of Hamiltonians belonging to the tenfold table are essentially Bogoliubov-de Gennes Hamiltonians [18, 20, 28] whose symbols are represented using Dirac matrices [9], such as:

$$\mathcal{H}(\boldsymbol{k}, \boldsymbol{r}) = \boldsymbol{h}(\boldsymbol{k}, \boldsymbol{r}) \cdot \boldsymbol{\gamma} \equiv \boldsymbol{h}_s \cdot \boldsymbol{\gamma}_s + \boldsymbol{h}_a \cdot \boldsymbol{\gamma}_a, \quad (2)$$

where $\boldsymbol{\gamma}$ represents a set of $(p + q + 1)$ matrices that both anti-commute and square to one, thereby forming the Clifford algebra $Cl_{q+1,p}$. For these matrices, the eigenvalues are $\pm|\boldsymbol{h}(\boldsymbol{k}, \boldsymbol{r})|$. Consequently, they are associated with an elliptic differential operator if the function $\boldsymbol{h}(\boldsymbol{k}, \boldsymbol{r})$ is zero at only a limited set of points. The components $\boldsymbol{h}_s$ and $\boldsymbol{h}_a$ are symmetric and anti-symmetric under momentum inversion, namely $\boldsymbol{h}_s(\boldsymbol{k}, \boldsymbol{r}) = \boldsymbol{h}_s(-\boldsymbol{k}, \boldsymbol{r})$ and $\boldsymbol{h}_a(\boldsymbol{k}, \boldsymbol{r}) = -\boldsymbol{h}_a(-\boldsymbol{k}, \boldsymbol{r})$. The dimensionality $w$ of the generators is determined by the algebraic structure of the Clifford algebra and is defined as

$$w = \begin{cases} 2^{\frac{p+q}{2}}, & p+q \text{ is even,} \\ 2^{\frac{p+q+1}{2}}, & p+q \text{ is odd.} \end{cases} \quad (3)$$

The symmetry class $s$ is determined by the relation $s = p - q \mod 8$ [8, 9], which provides an algebraic relation between the Dirac structure of the symbol and the tenfold classification. This indicates that any symbol described by the pair $(\delta, s)$, where $\delta = d - D$ is the reduced dimension, is topologically equivalent to a Dirac symbol with the identical pair. Moreover, there exists a continuous transformation between the two symbols that preserves both the symmetry class and $\delta$, indicating that no topological phase transition occurs throughout this process. Dirac symbols allow for the computation of a topological index $\nu_{\mathcal{H}}$, which captures the nonanalytic behavior of the symbol's eigenstates over phase space. In two dimensions, this index reduces to the Chern number, computed via the integral of a Berry curvature [25, 26]. Generally, in higher dimensions and in more complex symmetry classes, generalized Chern or winding numbers are applicable. While these invariants can be challenging to compute precisely, the Dirac structure of the symbol notably simplifies the expressions [9]. For $\nu_{\mathcal{H}}$ to have a finite (integer) value, the number of variables $(\boldsymbol{k}, \boldsymbol{r})$ needs to be equal to the number of components in $\boldsymbol{h}$,

$$d + D = p + q. \quad (4)$$

The topological index $\nu_{\mathcal{H}}$ is independent of the magnitude of $\boldsymbol{h}$, provided it does not vanish. This "flattening" condition, when applied to Bloch Hamiltonians, ensures the existence of a spectral gap. Here, it ensures the symbol is invertible, a condition for ellipticity. Equation (4) indicates that the sum $p + q$ and $d + D$ possess the same parity, similar to $s = p - q$ and $\delta = d - D$.

As a result, $s - \delta$ is necessarily even, and the condition $\delta - s = 2 \pmod 4$ excludes the $\mathbb{Z}$ topology. Therefore, only the case $\delta - s = 0 \pmod 4$ corresponds to integer topological invariants [9]. The introduction of the extra dimension $D$ modifies the count of symmetric $\boldsymbol{h}_s$ and anti-symmetric $\boldsymbol{h}_a$ fields, thereby affecting $(s, \delta)$. Consequently, $D$ enables the construction of $\mathbb{Z}$-topological phases.

Equation (4) enables us to express the symmetry class of a Hamiltonian characterized by a $\mathbb{Z}$ topology as $s = p - q = 2p - d - D$. Moreover, $p$ is defined as the number of anti-symmetric components relative to momentum in the symbol, which implies it is constrained by the spatial dimension $d$, thus $0 \le p \le d$. This implies that (recall $s$ is mod 8),

$$-d - D \le s \le d - D. \tag{5}$$

This condition restricts the accessible symmetry classes, depending on both the material's dimension and the defect introduced, itself being constrained by $d - 1$, meaning $0 \le D \le d - 1$. This allows for a thorough characterization of topological defects that generate a $\mathbb{Z}$ topology, determined by $d$ and the respective symmetry class $s$. Examples up to $d = 3$ are summarized in Table II.

| $d$ | $s$ | $D$ | Type of Defect |
|---|---|---|---|
| 1 | 1 | 0 | Point |
| 2 | 2, 6 | 0 | Line |
| 2 | 1, 5 | 1 | Point |
| 3 | 3, 7 | 0 | Surface |
| 3 | 2, 6 | 1 | Line |
| 3 | 1, 5 | 2 | Point |

TABLE II. Possible topological classes based on spatial dimension $d$ and defect dimension $D$.

We are now ready to outline our method for designing topological materials as needed. Initially, we have a non-topological material defined by $(d, s)$. By introducing a defect, we transform it into one of the topological symmetry classes within the same spatial dimension, as shown in Table II. To illustrate this idea, consider a material in symmetry class AI in three dimensions ($s = 0, d = 3$). Suppose we wish to introduce a defect that renders the material topological in symmetry class D i.e. with $s = 2$. According to Table II, this requires a line defect that contributes two additional symmetric terms to the symbol. This corresponds to coupling the system to a complex scalar field. A concrete example of such a transition occurs in a cubic lattice when a superconducting vortex line is introduced.

This approach enables transitions between a non-topological initial Hamiltonian $H_i$ to a desired topological Hamiltonian $H_f$ by introducing defect fields, all while maintaining the spatial dimension $d$. We refer to this as "navigating the tenfold table". Table III presents several examples from the literature, highlighting each by its defect type and illustrating the associated transitions in both reduced dimension $\delta$ and symmetry class $s$. The challenging aspect involves pinpointing the appropriate defect to introduce into the Hamiltonian, which will subsequently be converted into the correct field within the symbol that triggers a transition into a topological class. We now focus on a special family of materials that we label "chiral materials satisfying the Nielsen-Ninomiya fermion doubling theorem" [32, 33]. These materials host pairs of Dirac points and encompass various structures such as monolayer and multilayer graphene [34, 35], brickwall [36, 37], Kagome [38, 39], and Mielke lattices [40]. We offer a systematic approach to identify defects that induce transitions from BDI to BDI or BDI to CII, as detailed in Table III. Chiral Hamiltonians are of the form $H_C = \begin{pmatrix} 0 & \mathcal{Q} \\ \mathcal{Q}^\dagger & 0 \end{pmatrix}$, where the index of elliptic operators $\mathcal{Q}$ plays a role.

Consider an initial Hamiltonian $H_i$ representing a two-dimensional system within the class $BDI$ ($s = 1$), enjoying both T and P symmetries with $T^2 = P^2 = +1$. According to Table I and equation (4), this system falls under the non-topological category. The most general symbol for such a Hamiltonian is given by $\mathcal{H}_i(\boldsymbol{k}) = h_1(\boldsymbol{k})\sigma_x + h_2(\boldsymbol{k})\sigma_y$, where $h_1(\boldsymbol{k})$ is a symmetric function, and $h_2(\boldsymbol{k})$ an antisymmetric function w.r.t. $\boldsymbol{k}$. We present a systematic method, using point defects, to reach a topological class. A point defect in two dimensions relates to $D = 1$, which leads to a transition characterized by $\delta = d - D = 1$ within the symmetry classes $BDI$ or $CII$, depending on $\boldsymbol{h}(\boldsymbol{k})$. The construction proceeds by analyzing the spectrum of $\mathcal{H}_i(\boldsymbol{k})$, which is given by $\pm|\boldsymbol{h}|$. We examine the points where the spectrum vanishes (if any); by definition, these are Dirac points. The Nielsen–Ninomiya theorem [32, 33] states that these points must exist as pairs of non-equivalent points with opposite chiralities, meaning that the number of Dirac points must be even. A local perturbation, such as a point defect, couples long-wavelength modes at the same energy and therefore connects the two Dirac points. The topology of the material can be described through the low-energy expansion of the symbol near these points. The unperturbed symbol near the Dirac points takes the general form $\mathcal{H}_D(\boldsymbol{k}) = \tilde{h}_1(\boldsymbol{k})\sigma_x \otimes \tau_z + \tilde{h}_2(\boldsymbol{k})\sigma_y \otimes \boldsymbol{1}$, where in $\sigma_i \otimes \tau_j$, the Pauli matrix $\sigma_i$ refers to the chiral degrees of freedom and $\tau_j$ to the valley degrees of freedom. We aim to add a defect that leads to a complex scalar field so that the symbol now becomes:

$$\begin{aligned} \mathcal{H}_f(\boldsymbol{k}) = &\tilde{h}_1(\boldsymbol{k})\sigma_x \otimes \tau_z + \tilde{h}_2(\boldsymbol{k})\sigma_y \otimes \boldsymbol{1} \\ &+ \phi_1(\boldsymbol{r})\sigma_x \otimes \tau_x + \phi_2(\boldsymbol{r})\sigma_x \otimes \tau_y. \end{aligned} \tag{6}$$

The symmetry class is determined by $\tilde{\boldsymbol{h}}(\boldsymbol{k})$. In the simplest case of linear Dirac points, $\tilde{\boldsymbol{h}}$ is antisymmetric in $\boldsymbol{k}$, so that $s = p - q = 2 - 1 = 1$, i.e., class $BDI$. For a more complex case with quadratic Dirac points, $\tilde{\boldsymbol{h}}$ is symmetric in $\boldsymbol{k}$, leading to $s = p - q = 0 - 3 = -3 = 5 \pmod 8$, i.e., class $CII$. To create this type of defect, we start with a material that is translationally invariant and possesses Dirac points. We then expand its symbol in the vicinity of these points and introduce a field that fulfills (6). This defect maintains chiral symmetry and corresponds to a complex scalar field. The corresponding potential should possess arbitrary radial characteristics, connect the chiral degrees of freedom, and show angular symmetry around the defect. The last condition is less intuitive and is presented in the supplementary material. These conditions are formulated in the basis of the Dirac points; if the Hamiltonian is not originally written in that basis, the conditions must be modified accordingly. This construction guarantees the realization of a topological material in class $BDI$ or $CII$. We next demonstrate an example of each scenario.

| Material | $\delta \to \delta'$ | $s \to s'$ |
|---|---|---|
| SSH with a domain wall | $1 \to 1$ | BDI $\to$ BDI |
| Kitaev chain with domain walls | $1 \to 1$ | BDI $\to$ D |
| 2D p-wave superconductor on a square lattice with a vortex | $2 \to 1$ | D $\to$ D |
| 3D p-wave superconductor on a cubic lattice with a vortex | $3 \to 2$ | D $\to$ D |
| 2D chiral materials satisfying NN theorem with a "vacancy" | $2 \to 1$ | BDI $\to$ BDI or BDI $\to$ CII |
| 2D chiral materials satisfying NN theorem with an "adatom" | $2 \to 1$ | BDI $\to$ AI |

TABLE III. Examples of materials that turn topological upon introducing a defect [8, 28–31]. Chiral materials in $d = 2$ abiding by the Nielsen-Ninomiya theorem underlie our general construction in the text. Additional information is provided in the supplementary material.

The honeycomb lattice offers a classic example of a bipartite $(A, B)$ structure, suitable for the application of the aforementioned approach. It explains numerous properties observed in a graphene flake. Each site consists of identical atoms with a Bloch Hamiltonian [34, 41]:

$$H_i = t \sum_{\boldsymbol{k}} f(\boldsymbol{k}) \, a_{\boldsymbol{k}}^\dagger b_{\boldsymbol{k}} + h.c. \tag{7}$$

where $f(\boldsymbol{k}) \equiv 1 + e^{i \boldsymbol{k} \cdot \boldsymbol{a}_1} + e^{i \boldsymbol{k} \cdot \boldsymbol{a}_2}$, taking into account the lattice vectors $\boldsymbol{a}_1 = \sqrt{3} a \, \hat{\boldsymbol{x}}$ and $\boldsymbol{a}_2 = \frac{\sqrt{3}}{2} a \, \hat{\boldsymbol{x}} + \frac{3}{2} a \, \hat{\boldsymbol{y}}$, where $a$ is the lattice constant. Here, $t$ represents the energy for hopping between neighboring sites, and the onsite energy is set to zero. The symbol of this Hamiltonian is $\mathcal{H}_i(\boldsymbol{k}) = h_1(\boldsymbol{k}) \, \sigma_x + h_2(\boldsymbol{k}) \, \sigma_y$. The targeted Hamiltonian is $H_f$, characterized by $s_f = 1$, $d = 2$, and $D = 1$. From the range of potential Hamiltonians, we select the initial $d = 2$ graphene flake and incorporate a defect field with $D = 1$, ensuring the system belongs to the symmetry class $BDI$ ($s = 1$). The dimensionality reduction $\delta = d - D$ is equal to 1, which places $H_f$ in the $\mathbb{Z}$ class identified by a topological index $\nu_{\mathcal{H}_f}$ namely, a winding number. Its symbol is given by (6) with $\tilde{h}_1(\boldsymbol{k}) = k_x$ and $\tilde{h}_2(\boldsymbol{k}) = k_y$. Inverting this symbol allows one to obtain the target Hamiltonian $H_f$ ($s_f = 1$, $d = 2$, $D = 1$) and to identify the nature of the needed defect field, here a vacancy, i.e. the removal of one site in the lattice [9, 42–53]. The winding number is,

$$\nu_{\mathcal{H}_f} = \frac{1}{2\pi} \int d\theta \, \frac{1}{\phi_1^2 + \phi_2^2} \begin{vmatrix} \phi_1 & \phi_2 \\ \partial_\theta \phi_1 & \partial_\theta \phi_2 \end{vmatrix} = \mp \int \frac{d\theta}{2\pi} = \mp 1, \tag{8}$$

where $\boldsymbol{\phi}(\boldsymbol{r}) = \phi_1 + i\phi_2 = |\phi(r)| e^{\mp i\theta}$ [9]. Namely, the winding of the phase of $\boldsymbol{\phi}(\boldsymbol{r})$, encircling the vacancy, where $-1$ (and $+1$) signify a vacancy in sublattice $A$ (or $B$), respectively.

We now examine a transition between two Hamiltonians from distinct symmetry classes. It is particularly intriguing as it permits the incorporation of spin degrees of freedom that were initially absent. Starting from the same non-topological Hamiltonian $H_i$ ($s_i = 1$, $d = 2$, $D = 0$), we now look for a target topological Hamiltonian $H_f$ ($s_f = 5$, $d = 2$, $D = 1$). Equation (4) results in ($p = 0, q = 3$), with $w = 4$. This indicates the existence of four symmetric ($4 \times 4$) $\boldsymbol{\gamma}$ matrices. These matrices represent symmetric combinations of $\boldsymbol{k} = (k_x, k_y)$ and two real fields, $(\phi_1(\boldsymbol{r}), \phi_2(\boldsymbol{r}))$, specifically,

$$\mathcal{H}_{AB}(\boldsymbol{k}, \boldsymbol{r}) = \left( k_x^2 - k_y^2 \right) \sigma_z \otimes \tau_z + 2 k_x k_y \, \sigma_y \otimes \boldsymbol{1} \\ + \phi_1(\boldsymbol{r}) \, \sigma_x \otimes \tau_x + \phi_2(\boldsymbol{r}) \, \sigma_x \otimes \tau_y , \tag{9}$$

where $\boldsymbol{\phi}(\boldsymbol{r}) = \phi_1 + i\phi_2 = |\phi(r)| e^{\mp 2i\theta}$. Similarly to (8), the associated winding number now becomes $\nu_{\mathcal{H}_f} = \mp 4$. One factor of 2 originates from the quadratic energy dispersion at low energies, while the other will be detailed later on. The shift from the ($s = 1, \delta = d = 2$) class to the ($s = 5, \delta = 1, d = 2$) class highlights the interest of our method. Initially beginning with the $BDI$ class characterized by $T^2 = P^2 = +1$, our approach yields a Hamiltonian in the $CII$ class where $T^2 = P^2 = -1$. This corresponds to the emergence of a new spin-1/2-like degree of freedom within a spinless system. Therefore, by methodically incorporating engineered defects, we have the capability to generate effective spin-like degrees of freedom as desired.

An illustration of (9) is implemented in a tight bind-ing model for AB-stacked bilayer graphene (or brick wall lattice), where each layer includes one vacancy, both positioned directly one above the other (Fig. 2 (a)). In this setup, the phases $\theta$ related to the vacancies combine constructively, resulting in a phase of $2\theta$ and thus $\nu_{H_f} = \mp 4$. The underlying physical mechanism for this process is as follows: we begin with a $4 \times 4$ Bloch Hamiltonian whose band structure consists of four bands featuring two Dirac points (supplementary material). At zero energy, the Dirac points represent the crossing of the two middle bands. To obtain an effective symbol that describes the behavior near these Dirac points, we use Löwdin partitioning [54]. Through this method, it is shown that the symbol can be written as

$$\mathcal{H}_{\text{eff}}(\boldsymbol{k}) \propto \begin{pmatrix} 0 & f^2 \\ f^{*2} & 0 \end{pmatrix}, \qquad (10)$$

in the basis $(A_2, B_1)$. This Bloch Hamiltonian arises from an effective hopping interaction between $(A_2, B_1)$, depicted in Fig.1 (a).

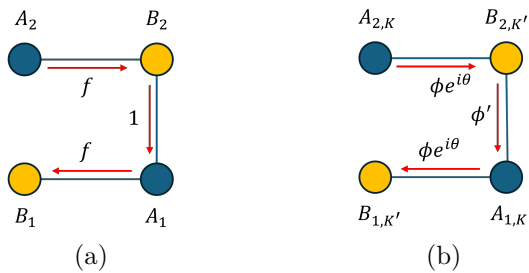

FIG. 1. (a) The red arrows illustrate the effective interaction between sublattices B in layer 1 and A in layer 2 as described by (10). (b) The red arrows depict the effective interaction caused by the vacancies field between sublattice B in valley $K'$ of layer 1 and sublattice A in valley $K$ of layer 2. The term $\phi'(r)$ is associated with the interaction between the vacancies. This is a localized field, and its exact form does not play a crucial role in the analysis.

The introduction of two vacancies, as shown in Fig. 2 (a), induces an interaction between the valleys of the two layers, as illustrated in Fig. 1 (b), resulting in the symbol in (9). This symbol features time-reversal symmetry, represented by $T = \sigma_z \otimes \tau_y K$, with the valley degree acting as an analogue of physical spin. Specifically, it leads to $T^2 = -1$. This suggests that vacancies not only induce valley coupling but also introduce a phase shift characterizing the valley transitions. This phase originates from the interlayer coupling described by (9). Because of the AB stacking arrangement, the two layers are oriented oppositely, resulting in structural asymmetry. This asymmetry leads to an extra phase factor in the valley coupling (see supplementary material for a detailed derivation).

Within this configuration of bilayer graphene with vacancies, two topological zero modes appear, each localized at a vacancy site. This phenomenon corroborates the ASIT theoretical prediction that connects the number $N_{ZM}$ of zero modes of a Hamiltonian with the winding number $\nu_{H_f}$, according to $N_{ZM} = |\text{Index } \mathcal{Q}| = \frac{1}{2}|\nu_{H_f}|$. This exemplifies the well-known bulk-edge correspondence, asserting that a nontrivial topological invariant in the bulk inherently leads to the emergence of protected edge states. Here, instead of manifesting at a physical boundary, the zero modes are localized at vacancy locations, serving as quasi-boundaries within the system. This observation broadens the idea of bulk-edge correspondence to encompass more than just traditional topological insulators, illustrating how lattice defects can be utilized to engineer and explore topological states.

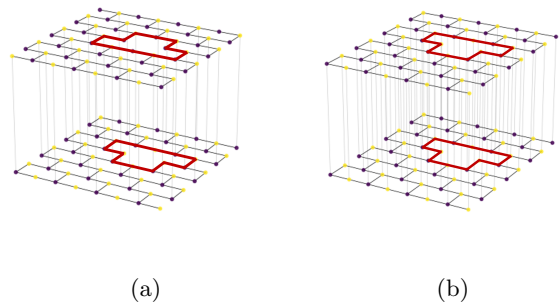

FIG. 2. A bilayer brick-wall lattice including a vacancy in each layer. These vacancies are aligned in two configurations: AB stacking (a) and AA stacking (b). The red outlines indicate the sites adjacent to the vacancy in each respective layer. In AB stacking, vacancies are positioned on opposite sublattices, resulting in the red contours being oriented oppositely.

A simpler topological phase transition can also take place between the BDI and AI symmetry classes. This change happens between the non-topological Hamiltonian $H_i$ ($s_i = 1, d = 2, D = 0$) and the non-topological Hamiltonian $H_f$ ($s_f = 0, d = 2, D = 1$). These relate to ($p = 1, q = 1$) and $w = 2$, denoted by the symbol:

$$\mathcal{H}(\boldsymbol{k}, \boldsymbol{r}) = h_1(\boldsymbol{k})\,\sigma_x + h_2(\boldsymbol{k})\,\sigma_y + m(\boldsymbol{r})\,\sigma_z, \qquad (11)$$

Here, $h_1(\boldsymbol{k})$ and $h_2(\boldsymbol{k})$ are respectively a symmetric and an anti-symmetric function with respect to $\boldsymbol{k}$. This symbol is accessible not only through the insertion of an adatom into graphene [9], but more fascinatingly by using AA stacking in bilayer graphene, as illustrated in Fig. 2 (b). In AA stacking, vacancies indeed break particle-hole symmetry.

In conclusion, we presented a systematic approach for exploring the tenfold table, enabling the transition from non-topological to topological phases via the introduction of defects. We demonstrated this technique through examples with monolayer and bilayer graphene having vacancies, which exhibit distinctive topological characteristics. We propose that this approach is largely applicable and can be expanded to systems beyond lattices by using the framework of quantum graphs [55–57]. Quantum graphs consist of vertices connected through edges and accommodate elliptic differential operators. Tight-binding models are just one particular example of quantum graphs which appear across a diverse range of topics, such as the physics associated with organic molecules, quantum chaotic billiards, and mesoscopic systems of electrons or photons [58] where our approach might also be applicable in designing topological systems. Another avenue to explore relates to quantum anomalies in gauge theories [59–61]. For continuum field theory, anomalies generally arise when a quantized Dirac field interacts with a gauge field and are commonly characterized by an index theorem within the ASIT framework. However, according to the Nielsen–Ninomiya theorem [32, 33], such anomalies cannot occur on a lattice because of fermion doubling: each fermion field is paired with a counterpart of opposite helicity, resulting in the anomaly's cancellation. For our situation, the vacancy field connects the two valleys (the two types of fermions), so its influence cannot be viewed as stemming from a typical gauge field. Instead, the valley coupling potential resulting from local defects reveals a unique mechanism. Consequently, our findings reveal novel physics that extends beyond the lattice-based interpretation of familiar gauge-field phenomena. A straightforward extension of this work involves constructing topological phases that facilitate quantum entanglement underlying more efficient quantum gates [10].

This research was funded by the Israel Science Foundation Grant No. 772/21 and the Pazy Foundation.

---

# Supplementary material for "Engineering Topological Materials"

Amit Goft and Eric Akkermans

*Department of Physics, Technion – Israel Institute of Technology, Haifa 3200003, Israel*

## I. SYMBOL DERIVATION FOR SELECTED EXAMPLES

In this section, we derive the symbol of the materials discussed in the main text and present additional examples to verify consistency with known results. Each example begins with a tight-binding Hamiltonian, which we transform using the discrete (Weyl) transform:

$$\mathcal{H}_{mn}(\boldsymbol{k}, \boldsymbol{r}) = \sum_j e^{-i\boldsymbol{k}\cdot\boldsymbol{R}(j)} \left\langle \boldsymbol{r} + \frac{\boldsymbol{R}(j)}{2} \Big|_m H \Big| \boldsymbol{r} - \frac{\boldsymbol{R}(j)}{2} \right\rangle_n . \tag{1}$$

Note that in Eq. (1), we redefined $\boldsymbol{R}(j) \to \frac{\boldsymbol{R}(j)}{2}$ for convenience, in contrast to the notation used in the main text. $m, n$ are matrix indices that correspond to additional degrees of freedom, such as sublattice, valley, etc.. We then recover the continuum limit and extract the principal symbol, eventually arriving at the Dirac form:

$$\mathcal{H}(\boldsymbol{k}, \boldsymbol{r}) = \boldsymbol{h}(\boldsymbol{k}, \boldsymbol{r}) \cdot \boldsymbol{\gamma} \equiv \boldsymbol{h}_s \cdot \boldsymbol{\gamma}_s + \boldsymbol{h}_a \cdot \boldsymbol{\gamma}_a. \tag{2}$$

We calculate topological invariants using [1]

$$\nu_{\mathcal{H}} = \frac{1}{S_{d+D}} \int_{S^{d+D}} d^d k \, d^D r \, J_{d,D}(\boldsymbol{h})$$

$$J_{d,D}(\boldsymbol{h}) = \begin{vmatrix} h_1 & h_2 & \cdots & h_{p+q+1} \\ \partial_1 h_1 & \partial_1 h_2 & \cdots & \partial_1 h_{p+q+1} \\ \vdots & \vdots & \ddots & \vdots \\ \partial_{d+D} h_1 & \partial_{d+D} h_2 & \cdots & \partial_{d+D} h_{p+q+1} \end{vmatrix}. \tag{3}$$

A summary of the resulting transitions appears in Table III of the main text.

### A. SSH chain

A well-known toy model for topological materials is the SSH chain [2, 3], whose Hamiltonian is given by

$$H = \sum_i t \, b_i^\dagger a_i - \sum_i t' \, a_{i+1}^\dagger b_i + \text{h.c.}. \tag{4}$$

Applying a Weyl transform to this Hamiltonian yields

$$
\begin{aligned}
\mathcal{H}_{AB}(k) &= \sum_{i,j} t \, e^{-ikaj} \, \langle x + \frac{j}{2}|_A \, a_i^\dagger b_i \, |x - \frac{j}{2}\rangle_B \\
&\quad - \sum_{i,j} t' \, e^{-ikaj} \, \langle x + \frac{j}{2}|_A \, a_{i+1}^\dagger b_i \, |x - \frac{j}{2}\rangle_B \\
&= t - t' e^{-ika}.
\end{aligned}
\tag{5}
$$

Since the symbol is Hermitian, we have $\mathcal{H}_{BA}(k) = \mathcal{H}_{AB}(k)^*$, and by definition $\mathcal{H}_{AA}(k) = \mathcal{H}_{BB}(k) = 0$. Therefore, the full symbol takes the matrix form

$$
\mathcal{H}(k) = \begin{pmatrix} 0 & t - t' e^{-ika} \\ t - t' e^{ika} & 0 \end{pmatrix} = f_1(k)\sigma_x + f_2(k)\sigma_y,
\tag{6}
$$

which is simply the Bloch Hamiltonian of the SSH model defined over the Brillouin zone $k \in [-\pi, \pi]$. Here, we define the real functions

$$
\begin{aligned}
f_1(k) &= t - t' \cos(ka), \\
f_2(k) &= -t' \sin(ka),
\end{aligned}
\tag{7}
$$

which satisfy the properties $f_1(-k) = f_1(k)$ and $f_2(-k) = -f_2(k)$, as expected. In the low-energy limit ($ka \ll 1$), we expand the Bloch Hamiltonian in powers of $k$, and define the mass parameter $m = t - t'$. This gives the continuum Dirac form

$$
\mathcal{H}(k) = k\sigma_y + m\sigma_x.
\tag{8}
$$

This Hamiltonian possesses all three fundamental symmetries: time-reversal symmetry (TRS), particle-hole symmetry (PHS), and chiral symmetry (CS), given by

$$
\begin{aligned}
T &= K, \\
P &= \sigma_z K, \\
C &= \sigma_z,
\end{aligned}
\tag{9}
$$

where $K$ denotes complex conjugation. These symmetries act on the symbol as follows:

$$
\begin{aligned}
T\mathcal{H}(k)T^{-1} &= K \left( f_1(k)\sigma_x + f_2(k)\sigma_y \right) K = f_1(k)\sigma_x - f_2(k)\sigma_y = \mathcal{H}(-k), \\
P\mathcal{H}(k)P^{-1} &= \sigma_z K \left( f_1(k)\sigma_x + f_2(k)\sigma_y \right) K \sigma_z = -f_1(k)\sigma_x + f_2(k)\sigma_y = -\mathcal{H}(-k), \\
C\mathcal{H}(k)C^{-1} &= \sigma_z \left( f_1(k)\sigma_x + f_2(k)\sigma_y \right) \sigma_z = -f_1(k)\sigma_x - f_2(k)\sigma_y = -\mathcal{H}(k).
\end{aligned}
\tag{10}
$$

Therefore, the SSH model belongs to symmetry class BDI. Since it is one-dimensional, it is characterized by a $\mathbb{Z}$-valued topological invariant, the winding number. To compute it, we define the complex function

$$
f_1(k) + if_2(k) = |f(k)|e^{i\phi(k)},
\tag{11}
$$

and use the formula (see Eq. (3)) for the winding number:

$$
\begin{aligned}
\nu_1 &= \frac{1}{2\pi} \int_{\text{BZ}} J(f_1, f_2) = \frac{1}{2\pi} \int_{\text{BZ}} \frac{1}{|f(k)|^2} \begin{vmatrix} f_1 & f_2 \\ \partial_k f_1 & \partial_k f_2 \end{vmatrix} dk \\
&= \frac{1}{2\pi} \int_{\text{BZ}} d\phi(k) = \begin{cases} 1 & t' > t, \\ 0 & t' < t. \end{cases}
\end{aligned}
\tag{12}
$$

Because the Bloch Hamiltonian is gapped, the Hamiltonian corresponds to an elliptic differential operator. Furthermore, the Brillouin zone is compact (a torus), so the Atiyah–Singer Index Theorem (ASIT) applies. We therefore expect the appearance of zero modes associated with the nonzero winding number. Such zero modes appear only in the presence of an edge. The simplest way to observe them is by considering a finite chain. In the trivial phase ($\nu_1 = 0$), no zero modes appear, while in the topological phase ($\nu_1 = 1$), two zero-energy modes are localized at the edges, as shown in Fig. 1. However, this result contains a subtlety: the index theorem predicts a single zero mode, in agreement with the winding number. Moreover, since the number of sites in both sublattices is equal, the analytical index of the Dirac operator vanishes. The resolution to this apparent contradiction lies in the interpretation of each edge as a point defect, each hosting a single zero mode. This will become clearer when we analyze the model in the presence of a defect.

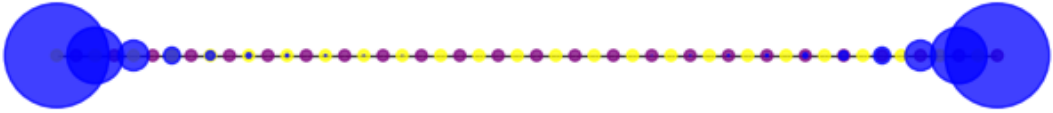

FIG. 1. One of the two zero modes localized on the edges of the lattice in the SSH chain.

## B. SSH chain with a point defect

We consider a defect characterized by a point in the lattice where the two hopping parameters $t$ and $t'$ switch, as shown in Fig. 2. This defect breaks translation invariance but preserves TR, PH

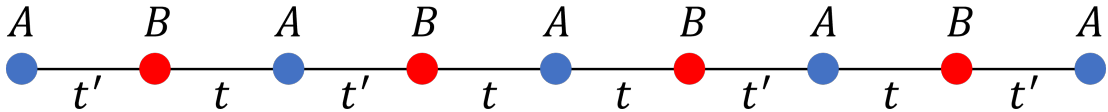

FIG. 2. SSH chain with a point defect, where the hopping parameters switch. Note that chiral symmetry is preserved.

and chiral symmetries. The Hamiltonian of this model is given by

$$H = \sum_{i \leq N} t'\, b_i^\dagger a_i - \sum_{i \leq N} t\, a_{i+1}^\dagger b_i + \sum_{i > N} t\, b_i^\dagger a_i - \sum_{i > N} t'\, a_{i+1}^\dagger b_i + \text{h.c.}$$
$$= \sum_i t_{1,i}\, b_i^\dagger a_i - \sum_i t_{2,i}\, a_{i+1}^\dagger b_i + \text{h.c.}, \tag{13}$$

where

$$t_{1,i} = \begin{cases} t' & i \leq N, \\ t & i > N, \end{cases}$$
$$t_{2,i} = \begin{cases} t & i \leq N, \\ t' & i > N. \end{cases} \tag{14}$$

Performing a Weyl transform on this Hamiltonian, we obtain

$$\mathcal{H}_{AB}(k,x) = \sum_{i,j} t_{1,i}\, e^{-ikaj}\, \langle x + \frac{j}{2} |_A\, a_i^\dagger b_i\, | x - \frac{j}{2} \rangle_B$$
$$+ \sum_{i,j} t_{2,i}\, e^{-ikaj}\, \langle x + \frac{j}{2} |_A\, a_{i+1}^\dagger b_i\, | x - \frac{j}{2} \rangle_B \tag{15}$$
$$= t_{1,x} - t_{2,x-\frac{1}{2}} e^{-ika}.$$

Thus, the Hamiltonian becomes

$$\mathcal{H}(k,x) = \begin{cases} \begin{pmatrix} 0 & t' - t e^{-ika} \\ t' - t e^{ika} & 0 \end{pmatrix} & x \leq N, \\[12pt] \begin{pmatrix} 0 & t - t' e^{-ika} \\ t - t' e^{ika} & 0 \end{pmatrix} & x > N. \end{cases} \tag{16}$$

In the low-energy limit ($ka \ll 1$) and in the continuum approximation, we expand the expression and define $m = t' - t$, resulting in (after setting $a = 1$ and omitting an overall minus sign)

$$\mathcal{H}(k,x) = k\sigma_y + m(x)\sigma_x, \tag{17}$$

where we have absorbed position dependence into the mass term, setting $m(x) = m \cdot \text{sgn}(x)$ and choosing $N = 0$. This expression reveals the defect as a domain wall in the continuum model. Importantly, this symbol still respects the three symmetries: TR, PH, and chiral symmetries, defined by

$$T = K,$$
$$P = \sigma_z K, \tag{18}$$
$$C = \sigma_z,$$

as can be verified by

$$T\mathcal{H}(k,x)T^{-1} = K\left(k\sigma_y + m(x)\sigma_x\right)K = -k\sigma_y + m(x)\sigma_x = \mathcal{H}(-k,x),$$
$$P\mathcal{H}(k,x)P^{-1} = \sigma_z K\left(k\sigma_y + m(x)\sigma_x\right)K\sigma_z = k\sigma_y - m(x)\sigma_x = -\mathcal{H}(-k,x),$$
$$C\mathcal{H}(k,x)C^{-1} = \sigma_z\left(k\sigma_y + m(x)\sigma_x\right)\sigma_z = -k\sigma_y - m(x)\sigma_x = -\mathcal{H}(k,x).$$

(19)

This confirms that the system belongs to symmetry class BDI, just like the SSH chain without a defect. To determine the topological invariant, we follow a similar procedure to that used in Eq. (12). However, here we integrate over the full phase space. The momentum space spans a circle $S^1$, while the defect splits real space into two regions, forming a sphere $S^0$ (i.e., two points) around the defect. Thus, $d + D = 1$. According to the tenfold classification, with $\delta = d - D = 1$ and $s = 1$, the model is again characterized by an integer topological invariant. Since $s$ is odd, this invariant is the one-dimensional winding number. Integrating over $S^0$ gives the difference in winding numbers across the defect:

$$\nu_1 = \frac{1}{2\pi}\int_{S^1\times S^0} dk\, dx\, J(\boldsymbol{h}[m(x)])$$
$$= \frac{1}{2\pi}\int_{S^1} dk\, J(\boldsymbol{h}[m]) - \frac{1}{2\pi}\int_{S^1} dk\, J(\boldsymbol{h}[-m]),$$

(20)

which simply gives the difference between the two possible winding numbers of the SSH model:

$$\nu_1 = \begin{cases} 1 & m > 0, \\ -1 & m < 0. \end{cases}$$

(21)

According to the ASIT, this topological invariant corresponds to a zero-energy mode localized at the domain wall, as shown in Fig. 3. This example illustrates a case in which introducing a defect does not change the codimension or the symmetry class, but it does modify the topological invariant of the system. In other words, compared to the uniform SSH chain, the system occupies the same cell in the periodic table, yet the value of the topological invariant changes.

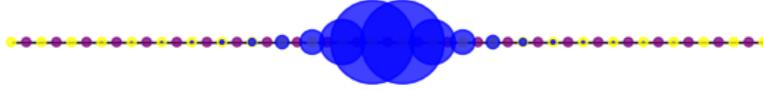

FIG. 3. Zero mode localized on the domain wall of an SSH chain.

## C. Graphene

To model graphene, we use a tight-binding approach on a honeycomb lattice. In this model, we consider only the nearest-neighbor hopping of spinless fermions, neglecting next-nearest-neighbor hoppings. With these assumptions, the Hamiltonian for pristine graphene is given by

$$H_0 = -t \sum_i \sum_{\alpha=0}^{2} a_i^\dagger b_{i+\alpha} + h.c.,$$

(22)

where $\alpha$ runs over nearest neighbors, and $a_i$ ($b_i$) is the annihilation operator at site $i$ in sublattice $A$ ($B$). The symbol for this Hamiltonian is the Bloch Hamiltonian

$$\begin{aligned}
\mathcal{H}_{AB}(\boldsymbol{k}, \boldsymbol{x}) &= -t \sum_j \sum_i \sum_{\alpha=0}^{2} e^{-i\boldsymbol{k}\cdot\boldsymbol{R}(j)} \left\langle x + \frac{j}{2} \right|_A a_i^\dagger b_{i+\alpha} \left| x - \frac{m}{2} \right\rangle_B \\
&= -t \sum_{\alpha=0}^{2} e^{-i\boldsymbol{k}\cdot\boldsymbol{R}(\alpha)} \\
&= -t \left( 1 + e^{i\boldsymbol{k}\cdot\boldsymbol{a_1}} + e^{i\boldsymbol{k}\cdot\boldsymbol{a_2}} \right),
\end{aligned}$$

(23)

with $\mathcal{H}_{AB}(\boldsymbol{k}, \boldsymbol{x}) = \mathcal{H}_{BA}^*(\boldsymbol{k}, \boldsymbol{x})$ and $\mathcal{H}_{AA}(\boldsymbol{k}, \boldsymbol{x}) = \mathcal{H}_{BB}(\boldsymbol{k}, \boldsymbol{x}) = 0$. Overall, we have

$$\mathcal{H}(\boldsymbol{k}, \boldsymbol{x}) = -t \begin{pmatrix} 0 & 1 + e^{i\boldsymbol{k}\cdot\boldsymbol{a_1}} + e^{i\boldsymbol{k}\cdot\boldsymbol{a_2}} \\ 1 + e^{-i\boldsymbol{k}\cdot\boldsymbol{a_1}} + e^{-i\boldsymbol{k}\cdot\boldsymbol{a_2}} & 0 \end{pmatrix}.$$

(24)

This system exhibits the following symmetries:

$$\begin{aligned}
T &= K, \\
P &= \sigma_z K, \\
C &= \sigma_z,
\end{aligned}$$

(25)

indicating that it belongs to symmetry class BDI. This symbol has $p = 1$, $q = 0$, and is associated with the Clifford algebra $Cl_{1,1}$. Unlike the SSH chain, graphene is two-dimensional, and hence it is not topological according to the tenfold classification. The two bands of this model intersect at zero energy, at six points commonly referred to as Dirac points. These points are named due to the linear spectrum, $E \propto \pm|k|$, around them. The Dirac points are divided into two equivalence sets, each containing three points. A common approach is to express (24) as a $4 \times 4$ matrix expanded around the Dirac points, which takes the form

$$\mathcal{H}_0(\boldsymbol{k}) = \begin{pmatrix} 0 & f(K+k) & 0 & 0 \\ f^*(K+k) & 0 & 0 & 0 \\ 0 & 0 & 0 & f(K'+k) \\ 0 & 0 & f^*(K'+k) & 0 \end{pmatrix},$$

(26)

where $f(\boldsymbol{k}) = 1 + e^{i\boldsymbol{k}\cdot\boldsymbol{a_1}} + e^{i\boldsymbol{k}\cdot\boldsymbol{a_2}}$. The regions in momentum space surrounding each Dirac point are referred to as "valleys." When representing the Bloch Hamiltonian as a $4 \times 4$ matrix, we explicitly include the valley degree of freedom. This representation is inherently redundant; however, as with any degenerate spectrum, the introduction of a small perturbation can couple degenerate states. In this context, perturbations can couple the two valleys.

Next, we analyze two types of defects in graphene that couple the valleys: a vacancy, and an adatom. We demonstrate that a vacancy is topological defect, meaning it shift pristine graphene into a topological class with a reduced effective dimension. In contrast, an adatom shifts graphene into a non-topological class.

### D. Graphene with a vacancy

A vacancy in graphene [1, 4–16] consists in the removal of a single neutral carbon atom from the lattice. In a tight binding description, it requires more careful consideration than pristine graphene. The vacancy disrupts the system's translational symmetry and couples the two valleys, breaking an additional unitary symmetry (parity symmetry). Hence, before calculating the Weyl transform of the Hamiltonian, we must write it explicitly, including both sublattice and valley degrees of freedom, to account for this coupling. The Hamiltonian of graphene with a vacancy placed at the origin, written in second quantization, is

$$H_V = -t \sum_i \sum_{\alpha=0}^{2} a_i^\dagger b_{i+\alpha} + t \sum_{\alpha=0}^{2} a_0^\dagger b_{0+\alpha} + h.c. \equiv H_0 + V, \tag{27}$$

where $H_0$ corresponds to pristine graphene, and $V$ denotes the vacancy potential. To account for the valley degree of freedom, we perform the following transformation on each creation and annihilation operator [15]:

$$\begin{aligned}
a_i &= \sum_k e^{-i\boldsymbol{k}\cdot\boldsymbol{R}(i)} a_k \\
&\equiv \sum_{k\in K} e^{-i\boldsymbol{k}\cdot\boldsymbol{R}(i)} a_k + \sum_{k\in K'} e^{-i\boldsymbol{k}\cdot\boldsymbol{R}(i)} a_k \\
&\equiv a_i^K + a_i^{K'},
\end{aligned} \tag{28}$$

since $k \in K(K')$ includes all points in the BZ in the vicinity of $K(K')$. This transformation requires to enlarge the Hamiltonian so as to include the valley degree of freedom. Note that states created by $a_i^{\dagger K}$ and $a_i^{\dagger K'}$ are orthogonal. Thus, we have

$$\begin{aligned}
H_0 &= -t \sum_i \sum_{\alpha=0}^{2} a_i^\dagger b_{i+\alpha} + h.c. \\
&= -t \sum_i \sum_{\alpha=0}^{2} \left( a_i^{\dagger K} b_{i+\alpha}^K + a_i^{\dagger K'} b_{i+\alpha}^{K'} \right) + h.c.,
\end{aligned} \tag{29}$$

where valley coupling terms vanish for pristine graphene. The Weyl transform of $H_0$ yields the expected symbol (Bloch Hamiltonian):

$$\mathcal{H}_0\left(\boldsymbol{k}\right) = \begin{pmatrix} 0 & 0 & f(K+\boldsymbol{k}) & 0 \\ 0 & 0 & 0 & f(K'+\boldsymbol{k}) \\ f^*(K+\boldsymbol{k}) & 0 & 0 & 0 \\ 0 & f^*(K'+\boldsymbol{k}) & 0 & 0 \end{pmatrix}, \tag{30}$$

where $f(\boldsymbol{k}) = 1 + e^{i\boldsymbol{k}\cdot\boldsymbol{a_1}} + e^{i\boldsymbol{k}\cdot\boldsymbol{a_2}}$. Unlike (26), here the symbol is written in the basis $(A, K; A, K'; B, K; B, K')$, e.g., the first row and the third column correspond to $\mathcal{H}_0\left(\boldsymbol{k}\right)_{A,K;B,K} = f\left(K+\boldsymbol{k}\right)$. Under the trans-formation (28), the vacancy potential takes the form

$$V = t \sum_{\alpha=0}^{2} \left( a_0^{\dagger K} + a_0^{\dagger K'} \right) \left( b_{0+\alpha}^{K} + b_{0+\alpha}^{K'} \right) + h.c. \tag{31}$$

The Weyl transform of this potential is

$$
\begin{aligned}
\mathcal{V}(\boldsymbol{k}, \boldsymbol{r})_{A,K;B,K'} &= \sum_{\alpha, \boldsymbol{r'}} e^{-i\boldsymbol{k}\cdot\boldsymbol{r'}} \langle \boldsymbol{r} + \frac{\boldsymbol{r'}}{2} | a_0^{\dagger K} b_{0+\alpha}^{K'} | \boldsymbol{r} - \frac{\boldsymbol{r'}}{2} \rangle \\
&= \sum_{\alpha} e^{i\boldsymbol{k}\cdot\boldsymbol{\alpha}} \delta_{\boldsymbol{r}, \frac{\boldsymbol{\alpha}}{2}}.
\end{aligned}
\tag{32}
$$

Next, we derive the principal symbol of $\mathcal{H}_V = \mathcal{H}_0 + \mathcal{V}$. Recall that is done by first taking the continuum limit and then considering only the highest powers in $\boldsymbol{k}$. The continuum limit is defined as $\boldsymbol{k} \cdot \boldsymbol{a} \leq 1$ and $\frac{r}{a} \geq 1$, i.e., we expand around the Dirac points ($\boldsymbol{k} \cdot \boldsymbol{a} \leq 1$) and assume the defect is localized in space ($\frac{r}{a} \geq 1$).

$$\mathcal{V}(\boldsymbol{k}, \boldsymbol{r})_{A,K;B,K'} \approx \sum_{\alpha} \left( e^{i\boldsymbol{K}\cdot\boldsymbol{\alpha}} + i\boldsymbol{k}\cdot\boldsymbol{\alpha} \right) \left( \delta(\boldsymbol{r}) + \frac{\alpha}{2}\partial_\alpha \delta(\boldsymbol{r}) \right). \tag{33}$$

The leading term in (33) is

$$\sum_{\alpha} e^{i\boldsymbol{K}\cdot\boldsymbol{\alpha}} \delta(\boldsymbol{r}) = 0. \tag{34}$$

To lowest order in $\boldsymbol{k} \cdot \boldsymbol{a}$, we obtain

$$
\begin{aligned}
\mathcal{V}(\boldsymbol{k}, \boldsymbol{r})_{A,K;B,K'} &\approx \sum_{\alpha} e^{i\boldsymbol{K}\cdot\boldsymbol{\alpha}} \frac{\alpha}{2} \partial_\alpha \delta(\boldsymbol{r}) \\
&\propto (i\partial_x + \partial_y)\delta(\boldsymbol{r}) \\
&= e^{i\theta} \left( -i\partial_r + \frac{1}{r}\partial_\theta \right) \delta(\boldsymbol{r}) \\
&= -ie^{i\theta} \partial_r \delta(\boldsymbol{r}).
\end{aligned}
\tag{35}
$$

The last equality follows from the fact that $\delta(\boldsymbol{r})$ depends only on $r$ and not on $\theta$. In the continuum limit, the symbol of the vacancy potential takes the form of a derivative of a delta function. This reflects the fact that the potential is extremely localized near the boundary of the vacancy, that is, around the bonds of the removed atom. As a result, the symbol of the potential vanishes both at the center of the vacancy and at locations far from the missing bonds. To make this behavior more explicit, consider approximating the Dirac delta function $\delta(\boldsymbol{r})$ using a Gaussian representation:

$$\delta(\boldsymbol{r}) \approx \lim_{\epsilon \to 0^+} \frac{1}{\sqrt{2\pi}\epsilon} e^{-\frac{r^2}{2\epsilon^2}}. \tag{36}$$

Taking the derivative, we find:

$$-i\partial_r\delta(\boldsymbol{r}) \approx \lim_{\epsilon\to 0^+} \frac{i\boldsymbol{r}}{\sqrt{2\pi}\epsilon^3}e^{-\frac{x^2}{2\epsilon^2}}. \tag{37}$$

This expression vanishes at the origin, reaches a maximum amplitude near the boundary of the vacancy (i.e., at $|\boldsymbol{r}| \sim \epsilon$), and decays exponentially away from it. Thus, the potential created by the vacancy has a characteristic length scale and specific spatial behavior. However, for topological considerations, the exact radial profile of the function is irrelevant; only its angular dependence matters. Consequently, we can replace $-i\partial_r\delta(r)$ with a general spatially localized function $\phi(\boldsymbol{r})$. We thus have

$$\mathcal{H}_V(\boldsymbol{k},\boldsymbol{r}) = \begin{pmatrix} 0 & 0 & Q & e^{i\theta}\phi(r) \\ 0 & 0 & e^{-i\theta}\phi(r) & -Q^\dagger \\ Q^\dagger & e^{i\theta}\phi(r) & 0 & 0 \\ e^{-i\theta}\phi(r) & -Q & 0 & 0 \end{pmatrix}, \tag{38}$$

where $Q = k_x - ik_y$. Using $\boldsymbol{\gamma}$ matrices, this symbol is

$$\mathcal{H}_V(\boldsymbol{k},\boldsymbol{r}) = k_x\sigma_x\otimes\tau_z + k_y\sigma_y\otimes\boldsymbol{1} + \phi_1(\boldsymbol{r})\sigma_x\otimes\tau_x + \phi_2(\boldsymbol{r})\sigma_x\otimes\tau_y, \tag{39}$$

where $\boldsymbol{\phi}(\boldsymbol{r}) \equiv \phi_1 + i\phi_2 = \phi(r)e^{\mp i\theta}$. Since the function $\boldsymbol{\phi}(\boldsymbol{r})$ is nonzero everywhere except at $\boldsymbol{r} = 0$, the symbol $\mathcal{H}_V$ vanishes only at a finite set of points—specifically, at $(\boldsymbol{k} = \boldsymbol{K}, \boldsymbol{r} = 0)$ and $(\boldsymbol{k} = \boldsymbol{K}', \boldsymbol{r} = 0)$. Therefore, $\mathcal{H}_V$ qualifies as an elliptic differential operator and can be incorporated into the tenfold classification. The vacancy preserves TRS, PHS, CS, placing the system in class BDI. This can be verified by counting the number of symmetric and antisymmetric coefficients with respect to $\boldsymbol{k}$ in the $\gamma$-matrices. In this case, there are two symmetric and two antisymmetric coefficients, giving $p = 2$ and $q+1 = 2$. Consequently, the symbol is described by the Clifford algebra $Cl_{2,2}$. Calculating the winding number (3) yields (after a simple integration over the momentum variables)

$$\nu_3 = \frac{1}{2\pi}\int d\theta\, \frac{1}{\phi_1^2+\phi_2^2}\begin{vmatrix} \phi_1 & \phi_2 \\ \partial_\theta\phi_1 & \partial_\theta\phi_2 \end{vmatrix} = \mp 1, \tag{40}$$

which is the winding of the phase of $\boldsymbol{\phi}(\boldsymbol{r})$ around the vacancy. The phase space in this case is described by the sphere $S^3$, a compact manifold. Additionally, since $H_V$ is an elliptic differential operator, the conditions of the ASIT are satisfied, and we expect the spectrum of $H_V$ to include zero modes. Specifically, a single vacancy creates an imbalance in the number of sublattice sites. For a finite system, this imbalance is expressed as:

$$\text{Index}(Q_V) = \text{DimKer}(Q_V) - \text{DimKer}(Q_V^\dagger) = N_A - N_B = V_B - V_A = \mp 1, \tag{41}$$

where $V_A$ and $V_B$ denote the number of vacancies in sublattices $A$ and $B$, respectively. As a result, a single zero mode is expected, corresponding to the presence of the vacancy. Fig. (4 a,b,c) illustrates several cases with varying boundary conditions. Each boundary condition produces a distinct spatial profile for the zero mode, a phenomenon explained by the Bulk-Edge correspondence. Topological zero modes are typically localized along the lattice edges. However, in this case, the zero mode can also localize at the vacancy site, suggesting that the vacancy serves as an effective edge for the lattice. This result naturally extends to systems with an arbitrary number of vacancies, $V_A$ and $V_B$.

The analytical index generalizes to Index$(Q_V) = V_B - V_A$, and we expect the topological invariant to reflect this value. Fig. (4 d) demonstrates a configuration with $V_A = 2$ and $V_B = 1$, which results in a single zero mode.

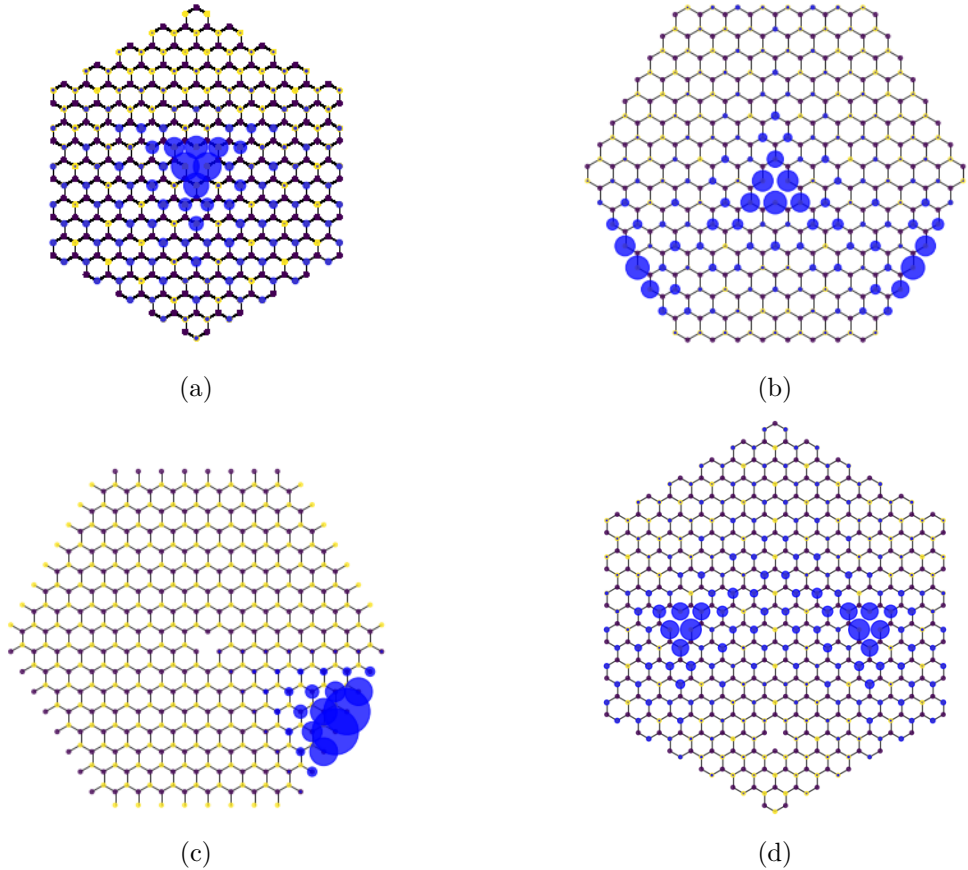

(a)

(b)

(c)

(d)

FIG. 4. Zero energy edge states from exact diagonalization of [27]. (a), (b), and (c) show armchair, zigzag, and bearded boundary conditions. (d) For three vacancies (two on $B$ and one on $A$), a single edge state appears on the majority $B$ sublattice. In each case, the zero mode is pinned to zero, highlighting its topological nature.

### E. Graphene with an adatom

We now examine a non-topological defect: the adatom [17, 18]. An adatom adds charge to one of the lattice sites, typically representing a hydrogen atom attached to a carbon atom. In a tight-binding framework, an adatom is modeled as an onsite energy at a specific lattice site, such

that

$$H_A = -t \sum_i \sum_{\delta=0}^{2} a_i^\dagger b_{i+\delta} + V_0 a_0^\dagger a_0 + h.c. \equiv H_0 + V_A, \tag{42}$$

where $H_0$ is the Hamiltonian for pristine graphene, and $V_A$ represents the adatom potential. Following the same approach as with a vacancy, we apply the transformation (28):

$$V_A = V_0 \left( a_0^{\dagger K} + a_0^{\dagger K'} \right) \left( a_0^K + a_0^{K'} \right) + h.c. \tag{43}$$

The Weyl transform in this case is straightforward, yielding in the continuum limit

$$\mathcal{V}_A(\boldsymbol{r}) = V_0 \delta(\boldsymbol{r}) \begin{pmatrix} 1 & 1 & 0 & 0 \\ 1 & 1 & 0 & 0 \\ 0 & 0 & 0 & 0 \\ 0 & 0 & 0 & 0 \end{pmatrix}, \tag{44}$$

so that $\mathcal{H}_A = \mathcal{H}_0 + \mathcal{V}_A$. Unlike a vacancy, $\mathcal{V}_A$ is real and corresponds to a single Dirac matrix in an equivalent Clifford algebra representation. It is also evident that time-reversal symmetry is preserved, while particle-hole and chiral symmetries are broken. An equivalent Clifford algebra representation requires three Dirac matrices, resulting in a 2x2 form:

$$\mathcal{H}_A(\boldsymbol{k}, \boldsymbol{r}) = h_1(\boldsymbol{k})\sigma_x + h_2(\boldsymbol{k})\sigma_y + m(\boldsymbol{r})\sigma_z, \tag{45}$$

where $h_1 - ih_2 = 1 + e^{-i\boldsymbol{k}\cdot\boldsymbol{a_1}} + e^{-i\boldsymbol{k}\cdot\boldsymbol{a_2}}$, and $m(\boldsymbol{r})$ corresponds to $\mathcal{V}_A$. This symbol is in symmetry class AI, with $p = q = 1$, indicating that it belongs to the Clifford algebra $Cl_{2,1}$. This gives $s = p - q = 0$ and a codimension $\delta = d - D = 2 - 1 = 1$. Therefore, graphene with an adatom lacks topological properties. A common misconception is the assumption that a vacancy and an adatom represent equivalent types of defects. Here, we highlighted the fundamental difference between them. It is true that, in the limit where the adatom potential becomes infinitely large, the system can exhibit behavior similar to that of a vacancy. This is because, in that limit, hopping to the adatom site is effectively prohibited, just as in the case of a vacancy. Indeed, numerical simulations confirm this correspondence. However, this limiting procedure must be approached with care. To fully reproduce vacancy-like behavior, not only must the adatom potential diverge, but the system size must also approach infinity. This is due to the fact that graphene with a vacancy has only one energy scale, namely $t$, while graphene with an adatom has two energy scales: $t$ and $V_0$. The system size introduces an additional energy scale of the order $tN$, where $N$ is the number of lattice sites. As a result, there is a competition between this energy scale and the adatom potential. The order in which the limits $N \to \infty$ and $V_0 \to \infty$ are taken can influence the physical behavior of the material. As demonstrated in [15], clear distinctions between vacancies and adatoms remain observable in physical measurements.

### F. Brick wall lattice

The next example is the brick wall lattice, shown in Fig. 5(a). This two-dimensional lattice resembles the honeycomb lattice in structure [19]. In fact, it is topologically equivalent to the honeycomb lattice, as it can be continuously deformed so that each site connects to three neighboring sites on the opposite sublattice. Like the honeycomb lattice, the brick wall lattice hosts a zero mode in the presence of a vacancy, as illustrated in Fig. 5(b). However, due to the absence of $C_3$ point group symmetry, the zero mode in the brick wall lattice decays anisotropically, with different rates in the $x$ and $y$ directions.

This equivalence between the honeycomb and brick wall lattices illustrates a broader principle: the system's topology is a property of the underlying graph, rather than the specific geometric layout of the lattice. In other words, topology depends on connectivity rather than spatial embedding. An advantage of the brick wall lattice is its rectangular structure, which simplifies numerical simulations compared to the hexagonal geometry of the honeycomb lattice. In our analysis, we impose periodic boundary conditions to eliminate boundary effects and isolate the influence of the vacancy.

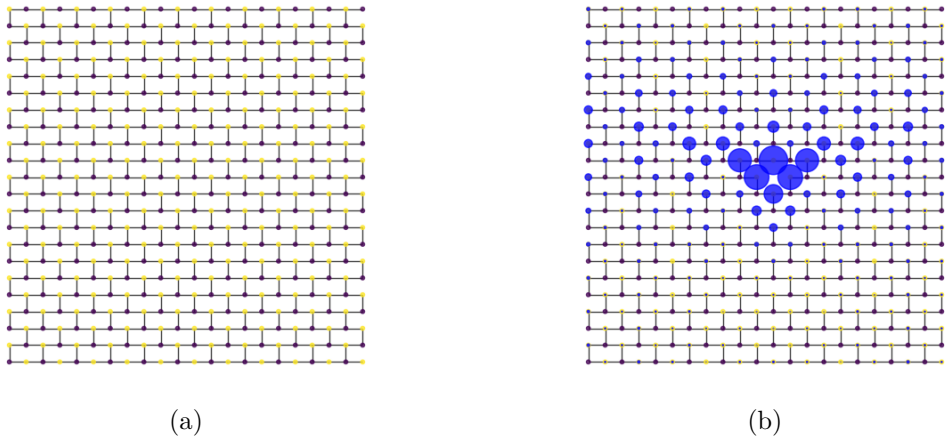

(a)                                    (b)

FIG. 5. (a) Brick wall lattice structure. (b) Brick wall lattice with a vacancy showing the presence of a zero mode.

### G. Bilayer brick wall lattice

When two layers of a brick wall lattice (or graphene) are attached, multiple stacking options are available. Different stacking configurations can either enhance or diminish the topological properties of a single layer. In general, stacking configurations fall into two categories. In the first type, the phases associated with the vacancies cancel each other out, resulting in a real scalar field, causing the material to lose its topological properties and shift it to a non-topological class in $\delta = 1$. In the second type, the phases are constructively added, creating a complex scalar field, which either maintains the topological class of the material or induces a shift to another topological class.

Here, we present two stacking examples: one that preserves the topological properties of monolayer graphene and another that enhances them, as shown in Fig. (2) in the main text.

## H. AB stacking

We start with a bilayer brick wall lattice with AB stacking, where each layer contains a vacancy, and the vacancies are aligned directly above one another, as shown in Fig. (**??**a). The tight-binding Hamiltonian for this model is given by

$$
\begin{aligned}
H = &-t \sum_i \sum_{\alpha=0}^{2} \sum_{l=1}^{2} a_{i,l}^\dagger b_{i+\alpha,l} - t \sum_i a_{i,1}^\dagger b_{i,2} \\
&+ t \sum_{\alpha=0}^{2} a_{0,1}^\dagger b_{0+\alpha,1} + t \sum_{\alpha=0}^{2} b_{0,2}^\dagger a_{0+\alpha,2} + t a_{0,1}^\dagger b_{0,2} + \text{h.c.} \\
\equiv & \, H_0 + V
\end{aligned}
\tag{46}
$$

Here, we consider AB stacking, as in bilayer graphene, where only sites in sublattice $A$ of the bottom layer are connected to sites in sublattice $B$ of the top layer.

In the basis $A_2, B_1, A_1, B_2$, the symbol of $H_0$ is

$$
\mathcal{H}_0(\boldsymbol{k}) = t \begin{pmatrix} 0 & 0 & 0 & f \\ 0 & 0 & f^* & 0 \\ 0 & f & 0 & 1 \\ f^* & 0 & 1 & 0 \end{pmatrix},
\tag{47}
$$

where $f = 1 + e^{i\boldsymbol{k}\cdot\boldsymbol{a_1}} + e^{i\boldsymbol{k}\cdot\boldsymbol{a_2}}$. Expanding this symbol around the Dirac points reveals two bands intersecting at the Dirac points with a parabolic dispersion (unlike the linear dispersion in monolayer graphene), while the other two bands do not intersect as illustrated in Fig. (6). To derive an effective Hamiltonian for the two intersecting bands, we use Löwdin partitioning [20]. This technique simplifies the treatment of Hamiltonians by partitioning them into subspaces. Consider a Hamiltonian $H$ represented in block form:

$$
H = \begin{pmatrix} H_{aa} & H_{ab} \\ H_{ba} & H_{bb} \end{pmatrix},
\tag{48}
$$

where $H_{aa}$ and $H_{bb}$ describe two subspaces with different energy scales, and $H_{ab}, H_{ba}$ are interaction terms between them. The goal is to derive an effective Hamiltonian for the subspace $a$ by eliminating subspace $b$. We solve the eigenvalue equation $H\psi = E\psi$, where $\psi$ is partitioned as:

$$
\psi = \begin{pmatrix} \psi_a \\ \psi_b \end{pmatrix}.
\tag{49}
$$

This leads to:

$$
\begin{aligned}
H_{aa}\psi_a + H_{ab}\psi_b &= E\psi_a, \\
H_{ba}\psi_a + H_{bb}\psi_b &= E\psi_b.
\end{aligned}
\tag{50}
$$

From the second equation, we solve for $\psi_b$:

$$
\psi_b = (E - H_{bb})^{-1} H_{ba}\psi_a.
\tag{51}
$$

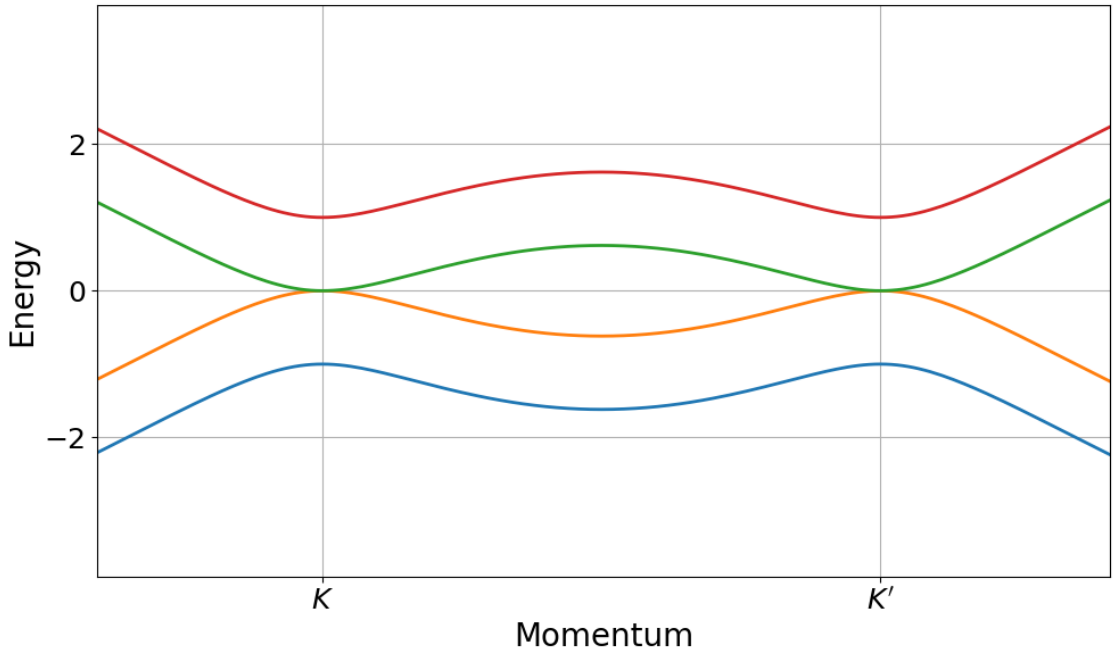

FIG. 6. The band spectrum of bilayer graphene features a pair of bands that cross at the Dirac points. In contrast to monolayer graphene, these Dirac points in bilayer graphene display a parabolic dispersion, meaning $E(\boldsymbol{k}) \propto |\boldsymbol{k}|^2$.

Assuming $(E - H_{bb})$ is invertible (valid since $H_{aa}$ and $H_{bb}$ differ in energy scale), substituting $\psi_b$ into the first equation yields the effective Hamiltonian:

$$H_{\text{eff}} = H_{aa} + H_{ab}(E - H_{bb})^{-1}H_{ba}. \tag{52}$$

Applying this method to AB-stacked bilayer graphene, we set $E = 0$, $H_{aa} = 0$, $H_{bb} = \begin{pmatrix} 0 & 1 \\ 1 & 0 \end{pmatrix}$, and $H_{ab} = H_{ba} = \begin{pmatrix} 0 & f \\ f^* & 0 \end{pmatrix}$. The effective Hamiltonian for the two intersecting bands becomes:

$$\mathcal{H}_{\text{eff}}(\boldsymbol{k}) \propto \begin{pmatrix} 0 & f \\ f^* & 0 \end{pmatrix} \begin{pmatrix} 0 & 1 \\ 1 & 0 \end{pmatrix} \begin{pmatrix} 0 & f \\ f^* & 0 \end{pmatrix} = \begin{pmatrix} 0 & f^2 \\ f^{*2} & 0 \end{pmatrix}, \tag{53}$$

which aligns with the parabolic nature of the Dirac points. Eq. (53) can be understood diagrammatically through Fig. 1 (a) in the main text. For the vacancy part, following the treatment of monolayer graphene, we examine the valley-coupling terms and incorporate interlayer coupling. With both valleys, the resulting $8 \times 8$ matrix (setting $t = 1$) is

$$\mathcal{H}\left(\boldsymbol{k}, \boldsymbol{r}\right) = \begin{pmatrix} 0 & 0 & 0 & f & 0 & 0 & 0 & \phi(r)e^{i\theta} \\ 0 & 0 & f^* & 0 & 0 & 0 & \phi(r)e^{i\theta} & 0 \\ 0 & f & 0 & 1 & 0 & \phi(r)e^{i\theta} & 0 & \phi'(r) \\ f^* & 0 & 1 & 0 & \phi(r)e^{i\theta} & 0 & \phi'(r) & 0 \\ 0 & 0 & 0 & \phi(r)e^{-i\theta} & 0 & 0 & 0 & f \\ 0 & 0 & \phi(r)e^{-i\theta} & 0 & 0 & 0 & f^* & 0 \\ 0 & \phi(r)e^{-i\theta} & 0 & \phi'(r) & 0 & f & 0 & 1 \\ \phi(r)e^{-i\theta} & 0 & \phi'(r) & 0 & f^* & 0 & 1 & 0 \end{pmatrix} \tag{54}$$

$$\equiv \begin{pmatrix} F\left(\boldsymbol{K}+\boldsymbol{k}\right) & \Phi\left(\boldsymbol{r}\right) \\ \Phi^\dagger\left(\boldsymbol{r}\right) & F\left(\boldsymbol{K}'+\boldsymbol{k}\right) \end{pmatrix},$$

where $\phi(r)e^{\pm i\theta}$ represents the field associated with a monolayer graphene vacancy, and $\theta$ is the angular coordinate encircling the vacancy location. The term $\phi'(r)$ corresponds to a field related to the coupling between the vacancies. It is a localized function, whose precise shape is not significant for the analysis.

Using Löwdin partitioning on $\Phi$, which has the same structure as $F$, we find that

$$\Phi_{\text{eff}}(\boldsymbol{r}) = \phi(r)\begin{pmatrix} 0 & e^{2i\theta} \\ e^{2i\theta} & 0 \end{pmatrix}, \tag{55}$$

where we redefined $\phi^2\left(r\right)\phi'\left(r\right) \to \phi\left(r\right)$. The phases of the vacancies add up in this case ($2\theta = \theta + \theta$), and the full symbol expanded around the Dirac points is

$$\mathcal{H}(\boldsymbol{k}, \boldsymbol{r}) = \left(k_x^2 - k_y^2\right)\sigma_x \otimes \tau_z + 2k_x k_y \sigma_y \otimes \mathbb{1} + \phi_1(r)\sigma_x \otimes \tau_x + \phi_2(r)\sigma_x \otimes \tau_y. \tag{56}$$

As mentioned in the main text, the vacancies induce two zero modes localized at the vacancy sites. This is illustrated in Fig. 7 (a) and (b).

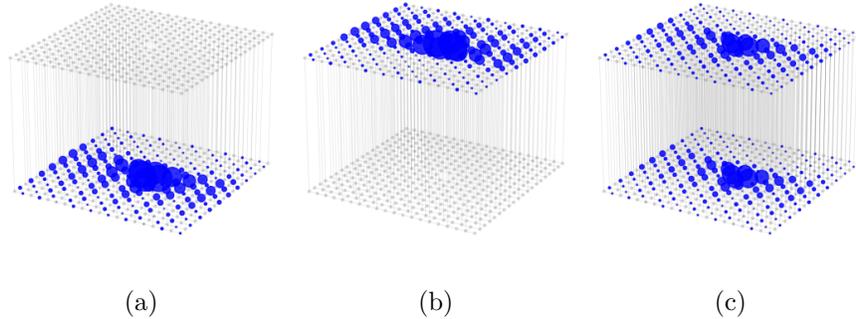

(a)　　　　　　(b)　　　　　　(c)

FIG. 7. (a) and (b) show two zero modes localized on the vacancies in bilayer brick wall lattice with AB stacking, consistent with the material's topological invariant. (c) shows a single zero mode in bilayer brick wall lattice with AA stacking at energy $+t$. A similar zero mode exist at energy $-t$.

## I. AA stacking

We now examine the bilayer brick wall lattice with AA stacking, where each layer contains a vacancy, and the vacancies are aligned directly above one another, as shown in Fig. 2 (b) in the main text. In AA stacking, there are two additional bonds within each unit cell. The tight-binding Hamiltonian for this model is given by:

$$
\begin{aligned}
H = &-t \sum_i \sum_{\alpha=0}^{2} \sum_{l=1}^{2} a_{i,l}^{\dagger} b_{i+\alpha,l} - t \sum_i \left( a_{i,1}^{\dagger} a_{i,2} + b_{i,1}^{\dagger} b_{i,2} \right) \\
&+ t \sum_{\alpha=0}^{2} a_{0,1}^{\dagger} b_{0+\alpha,1} + t \sum_{\alpha=0}^{2} a_{0,2}^{\dagger} b_{0+\alpha,2} \\
&+ t a_{0,1}^{\dagger} a_{0,2} + t b_{0,1}^{\dagger} b_{0,2} + \text{h.c.} \\
\equiv\ &H_0 + V.
\end{aligned}
\tag{57}
$$

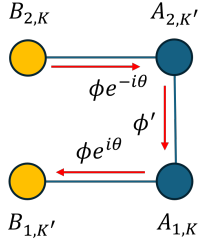

FIG. 8. Interactions in AA stacking bilayer graphene due to vacancies. The red contour represents the effective interaction between sublattice B in valley $K'$ of layer 1 and sublattice B in valley $K$ of layer 2 due to the vacancies field. Unlike AB stacking, the phases cancel each other out, leading to a an effective real scalar field and hence not a topological material.

Unlike AB stacking, this stacking configuration results in four Dirac points: two at energies $+t$ and two at energies $-t$ and none at zero energy. Consequently, at zero energy this material is non-topological. This is illustrated in Fig. (8) where the vacancies fields cancel each other out. This material, however, may exhibit topological features at energies $\pm t$. We shift the energy by $\pm t$ and expand around zero to explore this possibility. Shifting the energy by $-t$, we get:

$$
\mathcal{H}_0(\boldsymbol{k}) = t \begin{pmatrix} -1 & 1 & 0 & f \\ 1 & -1 & f & 0 \\ 0 & f^* & -1 & 1 \\ f^* & 0 & 1 & -1 \end{pmatrix},
\tag{58}
$$

where $f = 1 + e^{i\boldsymbol{k}\cdot\boldsymbol{a_1}} + e^{i\boldsymbol{k}\cdot\boldsymbol{a_2}}$, and this symbol is in the basis $A_1, A_2, B_2, B_1$. Applying a change of basis with the unitary matrix

$$U = \frac{1}{\sqrt{2}} \begin{pmatrix} 0 & 1 & 0 & 1 \\ 0 & 1 & 0 & -1 \\ 1 & 0 & 1 & 0 \\ 1 & 0 & -1 & 0 \end{pmatrix}, \tag{59}$$

we obtain

$$\mathcal{H}_0(\boldsymbol{k}) = t \begin{pmatrix} 0 & f^* & 0 & 0 \\ f & 0 & 0 & 0 \\ 0 & 0 & -2 & -f^* \\ 0 & 0 & -f & -2 \end{pmatrix}. \tag{60}$$

The blocks are decoupled, so Löwdin partitioning is unnecessary. This configuration results in Dirac points identical to those in monolayer graphene. We apply the same transformation to the defect component and obtain the full symbol (setting $t = 1$):

$$\mathcal{H}(\boldsymbol{k}, \boldsymbol{r}) = \begin{pmatrix} 0 & f^* & 0 & 0 & 0 & \phi(r)e^{-i\theta} & 0 & 0 \\ f & 0 & 0 & 0 & \phi(r)e^{i\theta} & 0 & 0 & 0 \\ 0 & 0 & -2 & -f^* & 0 & 0 & \phi'(r) & -\phi(r)e^{-i\theta} \\ 0 & 0 & -f & -2 & 0 & 0 & -\phi(r)e^{i\theta} & 0 \\ 0 & \phi(r)e^{-i\theta} & 0 & 0 & 0 & f^* & 0 & 0 \\ \phi(r)e^{i\theta} & 0 & 0 & 0 & f & 0 & 0 & 0 \\ 0 & 0 & \phi'(r) & -\phi(r)e^{-i\theta} & 0 & 0 & -2 & -f^* \\ 0 & 0 & -\phi(r)e^{i\theta} & 0 & 0 & 0 & -f & -2 \end{pmatrix}. \tag{61}$$

The symbol can be expressed in a block-diagonal form, since the first, second, fifth, and sixth rows and columns are decoupled from the remaining components. We focus on the blocks at energy $+t$, where expansion around the Dirac points yields

$$\mathcal{H}(\boldsymbol{k}, \boldsymbol{r}) = \begin{pmatrix} 0 & 0 & Q^* & e^{-i\theta}\phi(r) \\ 0 & 0 & e^{i\theta}\phi(r) & -Q \\ Q & e^{-i\theta}\phi(r) & 0 & 0 \\ e^{i\theta}\phi(r) & -Q^* & 0 & 0 \end{pmatrix}, \tag{62}$$

which matches the symbol for graphene with a vacancy (see [38] in a different basis. As such, it has the same winding number [40] and corresponds to a zero mode localized on the vacancies. Here, however, there are two zero modes, one at each energy level $\pm t$. In this case, the material remains in symmetry class BDI, with the transition involving only the spectral location of the topological features, from zero energy to $\pm t$. The zero mode is shown in Fig. [7] (c).